\documentclass[12pt]{article}
\usepackage{amsmath}
\usepackage{graphicx,psfrag,epsf}
\usepackage{enumerate}
\usepackage{natbib}
\setcitestyle{aysep={}}

\usepackage[english]{babel}
\usepackage{amsfonts}
\usepackage{amssymb}
\usepackage{hyperref}
\usepackage{tabularx}
\usepackage{xcolor}
\usepackage{booktabs,subcaption}

\newcommand{\blind}{1}
\newcommand{\githublink}{github.com/selbouhaddani/PO2PLS}

\makeatletter
\newcommand\newsubcommand[3]{\newcommand#1{#2\sc@sub{#3}}}
\def\sc@sub#1{\def\sc@thesub{#1}\@ifnextchar_{\sc@mergesubs}{_{\sc@thesub}}}
\def\sc@mergesubs_#1{_{\sc@thesub#1}}
\makeatother
\newsubcommand{\tto}{t}{\perp}
\newsubcommand{\uo}{u}{\perp}
\newsubcommand{\Tto}{T}{\perp}
\newsubcommand{\Uo}{U}{\perp}
\newsubcommand{\Wo}{W}{\perp}
\newsubcommand{\Co}{C}{\perp}

\newcommand{\tr}{\mathrm{tr}}
\renewcommand{\top}{\mathrm{T}}
\newcommand{\Cov}{\mathrm{Cov}}
\newcommand{\E}{\ensuremath{\mathbb{E}}}

\newcommand{\RR}{\ensuremath{\mathbb{R}}}
\newcommand{\orth}[1]{\ensuremath{\mathrm{orth}\left(#1\right)}}
\newtheorem{theorem}{Theorem}[section]
\newtheorem{AppThm}{Theorem}[section]

\newtheorem{definition}[theorem]{Definition}

\newcolumntype{Y}{>{\centering\arraybackslash}X}

\begin{document}

\def\spacingset#1{\renewcommand{\baselinestretch}%
{#1}\small\normalsize} \spacingset{1}


\if1\blind
{
  \title{\bf Statistical Integration of Heterogeneous Data with PO2PLS
}
  \author{Said el Bouhaddani\thanks{
    Said el Bouhaddani is an Assistant Professor (email: s.elbouhaddani@umcutrecht.nl) and Hae-Won Uh is an Associate Professor at the department of Data Science and Biostatistics, University Medical Center Utrecht, the Netherlands. Geurt Jongbloed is a Full Professor at the Delft institute of applied mathematics, Delft University of Technology, the Netherlands. Jeanine Houwing-Duistermaat is a Full Professor at the department of Statistical Sciences, University of Bologna, Italy, department of Statistics, University of Leeds, United Kingdom, and department of Data Science and Biostatistics, University Medical Center Utrecht, the Netherlands. 
    This work was supported by EU Horizon 2020 under Grant 721815 (IMforFUTURE); ERA-Net Erare-3 (MSA-Omics), EU IMI under Grant 116074 (BigData@Heart) and EU FP7-Health under Grant 305280 (MIMOmics). The authors acknowledge M. Harakalova, and M. Mokry, UMC Utrecht dept. of Cardiology, for providing data from the CVON-DOSIS HCM study; and C. Hayward and L. Klari\'c, university of Edinburgh MRC Institute of Genetics \& Molecular Medicine, for providing data from the CROATIA Korcula and Vis cohorts. }\hspace{.2cm}, 
    Hae-Won Uh, \\
    Department of Data science and Biostatistics, UMC Utrecht, Netherlands\\
    Geurt Jongbloed \\
    Delft Institute of Applied Mathematics, TU Delft, Netherlands\\
    and \\
    Jeanine Houwing-Duistermaat \\
    Department of Statistical Sciences, University of Bologna, Italy \\
}
  \maketitle
} \fi

\if0\blind
{
  \bigskip
  \bigskip
  \bigskip
  \begin{center}
    {\LARGE\bf Statistical Integration of Heterogeneous Data with PO2PLS}
\end{center}
  \medskip
} \fi

\bigskip
\newpage 
\begin{abstract}
The availability of multi-omics data has revolutionized the life sciences by creating avenues for integrated system-level approaches. Data integration links the information across datasets to better understand the underlying biological processes. However, high-dimensionality, correlations and heterogeneity pose statistical and computational challenges. We propose a general framework, probabilistic two-way partial least squares (PO2PLS), which addresses these challenges. PO2PLS models the relationship between two datasets using joint and data-specific latent variables. 
For maximum likelihood estimation of the parameters, we implement a fast EM algorithm and show that the estimator is asymptotically normally distributed.  
A global test for testing the relationship between two datasets is proposed, and its asymptotic distribution is derived.  Notably, several existing omics integration methods are special cases of PO2PLS.
Via extensive simulations, we show that PO2PLS performs better than alternatives in feature selection and prediction performance. In addition, the asymptotic distribution appears to hold when the sample size is sufficiently large.
We illustrate PO2PLS with two examples from commonly used study designs: a large population cohort and a small case-control study. Besides recovering known relationships, PO2PLS also identified novel findings. The methods are implemented in our R-package \textit{PO2PLS}.
Supplementary materials for this article are available online.
\end{abstract}

\noindent%
\vfill

\newpage
\spacingset{1.45} 

\section{Introduction}

Many studies collect multiple omics datasets to gather novel insights into various stages of biological processes:
genome-wide DNA markers reflecting the genetic code, transcriptomics and epigenetics, providing information on expressed and silenced genes, proteomics measuring the abundance of proteins. To link the information in 
these omics datasets, a joint integration approach is needed \citep{Richardson2016}. Several challenges exist: datasets are often high dimensional, measurements are highly correlated within and across datasets, and the presence of heterogeneity among datasets due to measuring different biological levels and using different technologies to measure them. Many machine learning methods have been proposed that address some of these challenges and are therefore increasingly popular \citep{Li2016a}. However, they neither provide statistical evidence for a relationship between the datasets nor identify relevant variables that contribute to this relationship. We propose a probabilistic latent variable modeling framework for inferring the relationship between two omics datasets $x$ and $y$. Our method reduces data dimensionality, captures correlations within and between sets of variables, addresses heterogeneity and performs statistical inference on the relation between $x$ and $y$. 

For our probabilistic approach, we propose to use multivariate normal distributions for $x$ and $y$. 
The correlation structure between and within $x$ and $y$ is modeled by joint and data-specific components, formed by linear combinations of the variables in $x$ and $y$. \citep{VanderKloet2016, Shu2020}. 
All parameters of the model are identifiable and estimated with maximum likelihood. To this end, a memory-efficient EM algorithm \citep{Meng1993} is implemented that can handle high dimensional data. We derive standard errors for the estimators and formulate a global test statistic for the null hypothesis of no relation between $x$ and $y$. For overparametrized models such as latent variable models, the regularity conditions under which maximum likelihood estimators are asymptotically normally distributed may not hold \citep{Sun2015}. We will derive the asymptotic distribution of our estimators as well as the distribution of our proposed test statistic; we apply the mathematical theory that investigates asymptotic properties of estimators based on minimizing a proper discrepancy function  \citep{Shapiro1983}. Finally, to deal with the size of the resulting asymptotic covariance matrix, which quadratically increases with the number of $x$ and $y$ variables, we develop an approximation that is very fast to compute, even in high dimensions. 
 


Various latent variable approaches are available, differing in models and estimation techniques. Instead of a maximum likelihood approach, algorithmic methods have been popular. These methods are often sequential, and the algorithm stops when the information in the datasets is sufficiently captured. Examples of algorithmic methods that only include joint parts are partial least squares (PLS) \citep{Wold1973} and canonical correlation analysis (CCA) \citep{Hotelling1936}. Methods also incorporating data specific parts are two-way orthogonal PLS (O2PLS) \citep{Trygg2003} and JIVE \citep{Lock2013}. JIVE is less flexible than O2PLS as it restricts the joint components of $x$ and $y$ to be exactly equal (details are given in Section \ref{par5:Genframew}). We have recently shown that when this assumption does not hold, convergence problems may arise and the performance of the estimators might be poor \citep{Bouhaddani2018b}. A shortcoming of algorithmic approaches is that standard errors are not available; hence a global test requires (computer-intensive) permutations to provide a p-value.

In contrast to algorithmic approaches, likelihood approaches provide a way to calculate standard errors. In addition, likelihood-based methods can assume a direction, i.e. $x$ influences $y$, which may lead to more efficient estimation of the true relation. An example is envelope regression \citep{Cook2015}, which fully models the covariance structure and is therefore not suited for high dimensional data. Alternatively, probabilistic PLS (PPLS) \citep{Bouhaddani2018} uses a simpler covariance structure with less parameters and is applicable to high dimensional datasets. In contrast to PPLS and envelope regression, SIFA \citep{Li2017} models specific components. However, just as JIVE, SIFA assumes the joint components to be exactly equal and might not perform well if this condition does not hold. Our novel data integration framework, probabilistic O2PLS (PO2PLS),  models joint and specific parts in $x$ and $y$. The models of SIFA and PPLS can be viewed as specific cases of our PO2PLS model.  

Nowadays, omics data are available in studies based on different designs, such as cross-sectional and follow-up population studies for common phenotypes, and case-control studies for rare diseases. We apply PO2PLS to data from a large cross-sectional population study and a small case-control study. For the population study, DNA markers ($p\approx10^5$) and glycomics ($q=20$) data are available for $N=885$ subjects \citep{Wahl2018}. This study has been part of genome-wide association studies, which test for associations between a marker and a glycan using single pair methods. Since glycans are highly correlated, these methods do not fully use the available information. We will perform a global test for association between the genetic markers and the glycan abundances, and assess which genes and glycans contribute most to this association. For the case-control study, epigenetics and transcriptomics data ($p,q\approx10^4$) are available for 23 subjects, of which 13 suffer from hypertrophic cardiomyopathy (HCM) and ten are healthy controls. Differential expression analyses are usually performed to infer significant relations between each pair of measurements. Instead, we globally test for an association between epigenetic activity and gene transcription. Genes in the joint components contributing to this association may play an important role in HCM.

The main contributions of this paper are threefold. We propose an EM algorithm to estimate the parameters of our PO2PLS model, which is computationally efficient and freely available on GitHub (\githublink) and will soon be released on CRAN. We formulate a global test to test the null hypothesis of no relationship between $x$ and $y$. We show the added value of our methods by applying them to omics datasets from two different studies. In Section \ref{sec5:M_and_E}, the PO2PLS model is formulated, and identifiability of the parameters is shown. Furthermore, maximum likelihood estimates are derived, and a global test of the relation between $x$ and $y$ is proposed. In Section \ref{sec5:SimuStudy}, the performance of PO2PLS is studied in a range of simulation scenarios. We focus on feature selection, prediction performance, type I error and power of the statistical test. In Section \ref{sec5:DataAnalysis}, PO2PLS is applied to the case studies to test and describe the relation between two sets of omics variables. We conclude with a discussion. 

\section{PO2PLS: model and estimation}
\label{sec5:M_and_E}
\subsection{The model}
Let $x$ and $y$ be two random row-vectors of size $p$ and $q$, respectively. 
In the PO2PLS model, both $x$ and $y$ are expressed in terms of a joint part, a specific part, and a noise part. The joint parts involve random vectors $t$ and $u$ of size $r$, with $r$ usually a small number. The specific parts involve independent random vectors $\tto$ and $\uo$ of size $r_x$ and $r_y$, respectively. The noise random vectors are denoted by $e$ ($p$-dimensional), $f$ ($q$-dimensional) and $h$ ($r$-dimensional). Here, $h$ represents heterogeneity in the joint parts, leading to differences between $t$ and $u$.
More precisely, the PO2PLS model for $x$ and $y$ is described by
\begin{equation}
\begin{split}
x = &\ tW^\top + \tto \Wo^\top + e, \quad y = \ uC^\top + \uo \Co^\top + f , \quad  u = tB + h \\
\end{split}
\label{eq4:PO2PLS}
\end{equation}
The parameter matrices $W$ ($p\times r$) and $C$ ($q\times r$) are called joint loadings. The matrices $\Wo$ ($p\times r_x$) and $\Co$ ($q\times r_y$) are referred to as data-specific loadings. 

The random vectors $e$ and $f$ are independent multivariate normally distributed random vectors, with zero mean and covariance matrices $\sigma_e^2 I_p$ and $\sigma_f^2 I_q$, respectively. Furthermore, $t$, $\tto$, $\uo$ and $h$ are zero mean multivariate normals, with diagonal covariance matrices $\Sigma_t$, $\Sigma_{\tto}$, $\Sigma_{\uo}$ and $\Sigma_{h}$, respectively. The covariance matrix of $u$ follows from \eqref{eq4:PO2PLS}: $\Sigma_u = B^\top \Sigma_t B + \Sigma_h$. Here, $B$ is a diagonal $r \times r$ matrix. 

{All parameters are collected in 
$\theta := [W, \Wo, C, \Co, B, \Sigma_t, \Sigma_{\tto}, \Sigma_{\uo}, \Sigma_{h}, \sigma^2_e, \sigma^2_f]$. It param\-eterizes the distribution of $(x,y) \sim \mathcal{N}(0, \Sigma_\theta)$ (the explicit expression for $\Sigma_\theta$ is given in the supplementary material). }

Note that the model for the relation between $u$ and $t$ is taken asymmetrically, as often a certain hierarchy is assumed for $x$ and $y$ \citep{Crick1970}. For instance, it is reasonable to assume that genetic variability induces glycomic variation, so a model for $u$ in terms of $t$ better reflects the underlying biology.

\paragraph{PO2PLS as a general data integration framework}
\label{par5:Genframew}
PO2PLS models the relationship between $x$ and $y$ through $t$ and $u$ as described in \eqref{eq4:PO2PLS}. It can be seen as a generalization of other models. Firstly, if the joint principal components (JPCs) are assumed to be exactly equal, i.e. $u=t$, the SIFA model is retrieved. In this case, $B=I$ and $\Sigma_h=0$, so $u$ and $t$ have the same scale and a correlation of one. However, datasets are typically heterogeneous, so the two sets of JPCs should represent different mechanisms (e.g. genetic versus glycomic pathways). Therefore, they may not be perfectly correlated or on the same scale. Also, assuming homogeneity of datasets can negatively affect estimation performance \citep{Bouhaddani2018b}.
Secondly, if additional to assuming $u=t$, the columns of the concatenated components $(W \Wo)$ and $(C \Co)$ are orthogonal, the JIVE model is recovered. In this case, combinations of features involved in the joint and specific parts have to be orthogonal, which is a strong restriction. 
Thirdly, the probabilistic PLS model is obtained by setting $\Sigma_{\tto}$ and $\Sigma_{\uo}$ to zero in \eqref{eq4:PO2PLS}. 

In the envelope regression (ER) model, the number of noise variance parameters to estimate is of order $O(p+q)$, whereas PPLS and PO2PLS introduce one $\sigma_e^2$ and $\sigma_f^2$ for $x$ and $y$, respectively. When $p$ or $q$ is larger than the sample size (i.e. a high dimensional setting) or the covariance matrix of $x$ or $y$ is singular, the ER estimator cannot be obtained due to singularity. 

From an estimation point of view, PO2PLS can be placed in the category of maximum likelihood estimators using EM (details about estimation is found below in Section \ref{subsec5:estim}). Other probabilistic approaches, such as ER, directly optimize the likelihood over Grassmann manifolds \citep{Cook2015}. Since this involves calculating the covariance of $(x,y)$, it is not feasible to use in high dimensions. Alternative approaches to maximum likelihood consider sequential algorithms to estimate joint and specific components. For example, the O2PLS estimator (see \cite{Trygg2003,Bouhaddani2016}) is as follows: first the covariance between $xW$ and $yC$ is optimized, then the covariance between $xW$ and $x-xWW^\top$ is optimized to get estimates for the specific parts, and finally after subtracting these parts, the covariance between $x^*W$ and $y^*C$ is optimized with the star indicating a deflation step. Our EM implementation of PO2PLS appears to be competitive with fast algorithmic approaches in terms of memory usage and is reasonably fast in high dimensional settings (see Section \ref{sec5:SimuStudy}).
In Table \ref{tab5:overview}, an overview is shown with several methods and their features.

\begin{table}[ht]
	\caption{\textbf{Features of several data integration methods.} An `X' indicates presence of a feature. All methods estimate a joint part. The first row indicates methods that also estimate specific components $W_\perp$ and $C_\perp$. The second row indicates methods that are based on a probability distribution for $x$ and $y$. For the next row, an X is placed if the method does not restrict the model to $u=t$. The last row indicates methods that can cope with data where $p,q$ are larger than the sample size. }
	\begin{tabularx}{\linewidth}{l*{8}{Y}}
		Properties & PLS & PPLS & ER & O2PLS & JIVE & SIFA & PO2PLS \\
		\toprule
		Specific 	&  &  & X & X & X & X & X \\ 
		Probab. 	&  & X & X &  &  & X & X \\ 
$u = tB + h$ 		& X &  & X & X &  &  & X \\
		High dim. 	& X & X &  & X & X & X & X \\
		\bottomrule
	\end{tabularx}
\label{tab5:overview}
\end{table}

\subsection{Identifiability of PO2PLS}
\label{subsec4:identif}
Linear latent variable models are typically unidentifiable due to rotation indeterminacy of the loading components. For example, given a rotation matrix $R$ such that $RR^\top = I$, the models $x=tW^\top$ and $x=(tR)(WR)^\top$ yield the same $x$ while $W$ and $WR$ are not the same. Note that if $\Cov(t)$ is diagonal with distinct elements, $\Cov(tR)$ is not diagonal unless $R$ is also diagonal. In PCA, the loading matrices are restricted to be semi-orthogonal, i.e. $W^\top W = I$, whereas in Factor analysis, the latent variables are standard normally distributed. However, these assumptions separately do not solve the rotation indeterminacy. In PO2PLS, identifiability can be obtained using similar assumptions, namely semi-orthogonal loading matrices and diagonal covariance matrices for the latent variables. 

The assumptions in PO2PLS are firstly, $W^\top W = C^\top C = I_r$, $\Wo^\top \Wo = I_{r_x}$ and $\Co^\top \Co = I_{r_y}$. Additionally, $[W \Wo]$ and $[C \Co]$ must not have linearly dependent columns. Note that the columns of $\Wo$ and $\Co$ do not have to be orthogonal to the columns of $W$ and $C$, respectively. Second, the diagonal elements of $B$ are restricted to be positive. This does not restrict the PO2PLS model, as $t_k b_k$ is equal to $-t_k b_k$ in distribution, for $k = 1,\ldots,r$. Finally, the sequence $(\sigma^2_{t_k}b_k)_{k=1}^r$ is assumed to be strictly decreasing in $k$. 
Regarding the number of components, we assume that $0 < r+r_x < p$ and $0 < r+r_y < q$, where $r$ is positive and both $r_x$ and $r_y$ are non-negative. 

Given these assumptions, the loading matrices are identified up to sign and the other parameters in $\theta$ are uniquely identified. The following Theorem makes this precise.
\begin{theorem}
Let $r$, $r_x$, $r_y$ and $\theta$ satisfy the above assumptions. Let $\Sigma_{\theta_1}$ and $\Sigma_{\theta_2}$ be the covariance matrices corresponding to PO2PLS parameters $\theta_1$ and $\theta_2$, and suppose $\Sigma_{\theta_1} = \Sigma_{\theta_2}$. Then $W_1 = W_2 \Delta_W$, $C_1 = C_2 \Delta_W$, $\Wo_1 = \Wo_2 \Delta_{\Wo}$, $\Co_1 = \Co_2 \Delta_{\Co}$ for diagonal orthogonal matrices $\Delta_W, \Delta_{\Wo}$ and $\Delta_{\Co}$, and all other parameters in $\theta_1$ and $\theta_2$ are equal. 
\end{theorem}
The proof is given in the supplementary material. 

\subsection{Maximum Likelihood Estimation of the parameters}
\label{subsec5:estim}
We propose maximum likelihood to estimate $\theta$. Contrary to the sequential O2PLS algorithm, the estimation is simultaneous over both joint and specific parts.
The log of the likelihood associated with the PO2PLS model \eqref{eq4:PO2PLS} is given by
\begin{equation}
L(\theta|x,y) = -\frac{1}{2} \left\{ (p+q) \log\left(2\pi\right) + \log\left|\Sigma_\theta\right| + (x,y)\Sigma_\theta^{-1}(x,y)^\top \right\}.
\end{equation}
Note that $L$ is a complicated and highly non-linear function of $\theta$, and its computation requires computing and storing covariance matrices of size $(p+q)^2$. If the latent variables $t$, $u$, $\tto$ and $\uo$ would be observable, maximizing the log-likelihood becomes analytically tractable and computationally feasible, even for large $p$ and $q$. However, the latent variables are not observable. In an EM algorithm \citep{Dempster1977}, predictions are calculated for these missing quantities, and iteratively, maximizers are obtained. Therefore, we propose an EM algorithm to obtain maximum likelihood estimates for $\theta$. 

Denote the complete data vector by $(x,y,t,u,\tto,\uo)$. For each current estimate $\theta'$, the EM algorithm considers the objective function
\begin{equation}
Q(\theta|x,y,\theta') := \E_{\theta'} \left[ \log f(x,y,t,u,\tto,\uo|\theta) | x,y \right].
\label{eq4:Qcompl}
\end{equation}
Here, the complete data likelihood can be written (with abuse of notation) as
\begin{equation}
f(x,y,t,u,\tto,\uo|\theta) = f(x|t,\tto)\,f(y|u,\uo)\,f(u|t)\,f(t)f(\tto)\,f(\uo).
\label{eq4:f_factors}
\end{equation}
These factors depend on distinct sets of parameters. For example $f(x|t,\tto)$ depends only on $W$, $\Wo$ and $\sigma_e^2$, yielding separate optimization problems.

The expectation step involves a conditional expectation of the complete data likelihood. Since $f$ in \eqref{eq4:Qcompl} is a multivariate normal density, this expectation can be written in terms of the first and second conditional moments of the latent variables $t$, $u$, $\tto$ and $\uo$ given $x$ and $y$. Focusing on the first factor in \eqref{eq4:f_factors}, the conditional expectation of $\log f(x|t,\tto)$ is given by
\begin{equation}
\begin{split}
-\frac{1}{2} \left\{ N p \log\left(2\pi\right) + N p \log\sigma_e^2 + \sigma_e^{-2}\tr\,\E_{\theta'}\left[||x-tW^\top-\tto \Wo^\top||_F^2\,| x,y \right] \right\}.
\end{split}
\label{eq4:logx}
\end{equation}
This expectation involves first and second conditional moments of the vector $(t, \tto)$ given $\theta'$, $x$ and $y$. These terms can be explicitly calculated and are given in the supplementary material.

In the maximization step, the function in \eqref{eq4:logx} is optimized over all semi-orthogonal matrices $W$ and $\Wo$. 
By introducing Lagrange multipliers $\Lambda_W$ and $\Lambda_{\Wo}$, maximizing \eqref{eq4:logx} over semi-orthogonal $W$ and $\Wo$ is then equivalent to minimizing the following objective function 
\begin{equation}
\E_{\theta'}\left[||x-tW^\top-\tto \Wo^\top||_F^2\,| x,y \right] + \Lambda_W \left(W^\top W - I_r\right) + \Lambda_{\Wo} \left(\Wo^\top \Wo - I_{r_x}\right).
\label{eq4:logxPen}
\end{equation}
Note that the objective function involves both $W$ and $\Wo$ and cannot be decoupled. Instead of numerical optimization, we consider a variant of EM that performs sequential optimization \citep{Meng1993}. First, \eqref{eq4:logxPen} is minimized over $W$, keeping $\Wo$ constant. Then we minimize over $\Wo$, keeping $W$ equal to its minimizer. Under standard conditions, this algorithm monotonically approaches a (local) maximum of the observed likelihood $L$ \citep{Meng1993}.

The above derivation is conditional on the dimensions of the latent spaces. Typically, the number of components $r$, $r_x$ and $r_y$ are unknown a priori. Strategies that can be used to select the number of PO2PLS components include cross-validation \citep{Geisser1993} and eigenvalue (scree) plots \citep{Mardia1979}.

The expectation and maximization step for the other parts in \eqref{eq4:f_factors} are calculated analogously (see the supplementary material). In this calculation, the orthogonalization operator is used to obtain semi-orthogonal loading matrices, defined as follows.
\begin{definition}
Let $A$ be a $p\times a$ full rank matrix with singular value decomposition $A = UDV^\top$. Let $R=VD$. Then we define the operator ${orth}: \RR^{p\times a} \to \RR^{p\times a}$ as $\orth{A} = A (R^\top)^{-1}$.
\end{definition}
Using this operator, the EM parameter updates are made explicit in Theorem \ref{th5:ECM} in the Appendix.

\subsection{Statistical inference: formulation of a global test}
\label{subsec5:asymp}
One of the challenges in data integration is to assess the statistical evidence for the relationship between $x$ and $y$. In our model, this relationship is represented by the equation $u = tB + h$ in \eqref{eq4:PO2PLS}. Thus the null hypothesis of no relationship corresponds with 
\begin{equation}
    H_0: B = 0 \qquad \mathrm{against} \qquad H_1: B \neq 0.
    \label{eq5:hypothesis} 
\end{equation}
To test this null hypothesis, we propose the following Wald-type test statistic,
\begin{equation}
    T_{B} = \hat{B} / \hat{SE}_{\hat{B}}.
    \label{eq5:T_B}
\end{equation}
We refer to \eqref{eq5:hypothesis} with \eqref{eq5:T_B} as the global test. To apply the global test statistic in practice, the asymptotic distribution of all parameters $\theta$, including $B$, needs to be derived. Since our model is overparameterized, standard maximum likelihood theory cannot readily be implemented. 

Under certain regularity conditions, consistency of the estimator $\hat{\theta}$ and its asymptotic distribution $\mathcal{N}(\theta, \Pi_\theta)$ follows from Shapiro's Proposition 4.2 \citep{Shapiro1986} applied to the PO2PLS model \eqref{eq4:PO2PLS}. Here, a suitable discrepancy function $F(S,\Sigma_\theta) = L(S) - L(\Sigma_\theta)$ with $S$ the sample covariance matrix of $(x,y)$ is used. Details and proofs are given in the supplement. 

Given the asymptotic covariance matrix $\Pi_\theta$, standard errors for the elements of $\hat{\theta}$ are obtained by calculating the square root of the diagonal elements of $\Pi_\theta$. 
An estimate of $\Pi_\theta$ is obtained from the inverse observed Fisher information matrix. In an EM algorithm, this matrix is given by \citep{Louis1982}:
\begin{equation}
\mathcal{I}_{\hat{\theta}} = \mathbb{E}\left[ B(\hat{\theta}) | X, Y \right] - \mathbb{E}\left[ S(\hat{\theta})S(\hat{\theta})^\top | X, Y \right].
\end{equation}
Here, $S(\hat{\theta}) = \nabla L(\hat{\theta})$ and $B(\theta) = -\nabla^2 L(\hat{\theta})$ are the gradient and negative of the second derivative of the log likelihood $L$, respectively, evaluated in $\hat{\theta}$. 
The derivation of the Fisher information matrix for the parameters of the PO2PLS model is given in the supplement. 

To obtain the standard errors for $\hat{B}$, the submatrix of $\mathcal{I}^{-1}_{\hat{\theta}}$ with respect to $B$ has to be calculated. However, this requires inverting a matrix of size $O((pqr)^2)$, which is computationally infeasible even for moderate $p$ and $q$. 
Under the assumptions that $\hat{B}$ and $\hat{\theta}/\hat{B}$ are asymptotically independent and $\hat{\Sigma_h}$ is non random, the observed Fisher information matrix $\mathcal{I}_{\hat{B}}$ and thus $SE_{\hat{B}}$ are given by the following formula, 
\begin{equation}
\mathcal{I}_{\hat{B}} = \hat{\Sigma}_h^{-1} \mathbb{E}[t^\top t|x,y] - \hat{\Sigma}_h^{-2} \mathbb{E}[(u-t\hat{B})^\top t t^\top  (u-t\hat{B})|x,y].
\label{eq5:IobsB}
\end{equation}
Details of the derivation of this formula are given in the supplementary material. Note that the first part on the right-hand side is the Fisher information matrix based on the general linear model, had $t$ and $u$ been observed. Standard errors for $\hat{B}$ are given by the square root of the diagonal elements of $\mathcal{I}^{-1}_{\hat{B}}$. Thus, to test the global hypothesis \eqref{eq5:hypothesis}, we apply our statistic $T_{B}$ and calculate the corresponding p-value. 

\section{Simulation study}
\label{sec5:SimuStudy}
We conduct a simulation study to evaluate the performance of PO2PLS in terms of feature selection, prediction and performance of our global test. Four metrics are considered: true positive rates, root mean squared error of the prediction, type I error and power. We compare PO2PLS to existing approaches PLS, O2PLS, PPLS and SIFA, covering algorithmic and probabilistic methods with and without specific parts (see Table \ref{tab5:overview}). We investigate robustness against model assumptions. Finally we assess computational efficiency.

For performance in feature selection and prediction ability, we consider combinations of small and large sample sizes ($N=100,1000$) and low and high dimensional data ($p=2000,10000$; $q=25,125$). We also include two proportions of noise relative to the total variation: in the `small noise proportion', we set the variance of $e$ and $f$ to be $40\%$ of the variance of $x$ and $y$. In the `large noise proportion', these values are $95\%$ and $5\%$ for $x$ and $y$, respectively. We set $B=I$ and $\Sigma_h = 0$ to comply with the SIFA assumptions, see Section \ref{par5:Genframew}. The impact of heterogeneity of joint parts is considered by increasing the joint residual variance $\Sigma_h$ from 0\% to 80\% of the total joint variance $\Sigma_u$. Finally, we set $r$, $r_x$, and $r_y$ to five components. These scenarios are commonly encountered in data analysis.

To assess the feature selection performance, we calculate the proportion of true top 25\% features among the estimated top 25\% (i.e True Positives Rate, TPR). We then average these proportions across components to obtain an aggregated measure. 
Predictive performance is measured by calculating the RMSEP, defined as the square root of $\mathbb{E}||y-\hat{y}||^2$ with $\hat{y}$ predicted from $x$. The RMSEP is calculated in both training and test data; the test data consist of $N=10^4$ independent samples generated from the same model as the training data. 

To evaluate the performance of the PO2PLS global test $T_B$ described in Section \ref{subsec5:asymp}, we 
first estimate the type I error for increasing sample size of $50$, $500$, $5000$, and $10000$. The dimension of $x$ is set to $20$. The number of simulation replicates here is 50000. Next, we consider increasing dimensionality, namely $p=20, 200, 2000$. The sample size is set to 500, and we replicate 2000 times. The type I error is calculated as the proportion of rejecting the null hypothesis $B=0$ at a 5\% level when simulating under this hypothesis.
Next, we estimate the power of $T_B$ in \eqref{eq5:T_B}. We compare four procedures, namely using the normal distribution for $T_B$ with calculated standard errors using our approximation of its covariance matrix, with standard errors obtained from parametric and from non-parametric bootstrapping, and with using the empirical distribution of $T_B$ via permutations. The proportion of false and true rejections are reported and compared for increasing $B$. We consider a sample size of $50$ and $500$, and dimensionality $p$ of $20$ and $200$. The number of bootstrap and permutation iterations is 250 and 500, respectively, and we repeat 500 times. In all three simulations, the dimension of $y$ is kept to $5$, the noise proportion is 50\%, and we set $r=2$, $r_x=1$, and $r_y=0$. 

Three additional simulation studies are carried out to study the robustness of PO2PLS against model deviations.
PO2PLS is applied to high dimensional simulated datasets from a selected case-control study design, mimicking the second data analysis in Section \ref{sec5:DataAnalysis}. We compare the error of predicting the outcome using the PO2PLS joint components with aforementioned alternatives. 
Then, we assess the impact of rank misspecification when fitting PO2PLS, by estimating too few components, and the impact of non-normality of the latent variables, using four commonly encountered distributions. 
Finally, we study the computational efficiency of the PO2PLS implementation, measured by the cpu time and memory demand of the EM algorithm. Details of these simulations and results are given in the supplementary material. 

\subsection{Simulation results}

\begin{figure}[ht]
	\includegraphics[width=\textwidth]{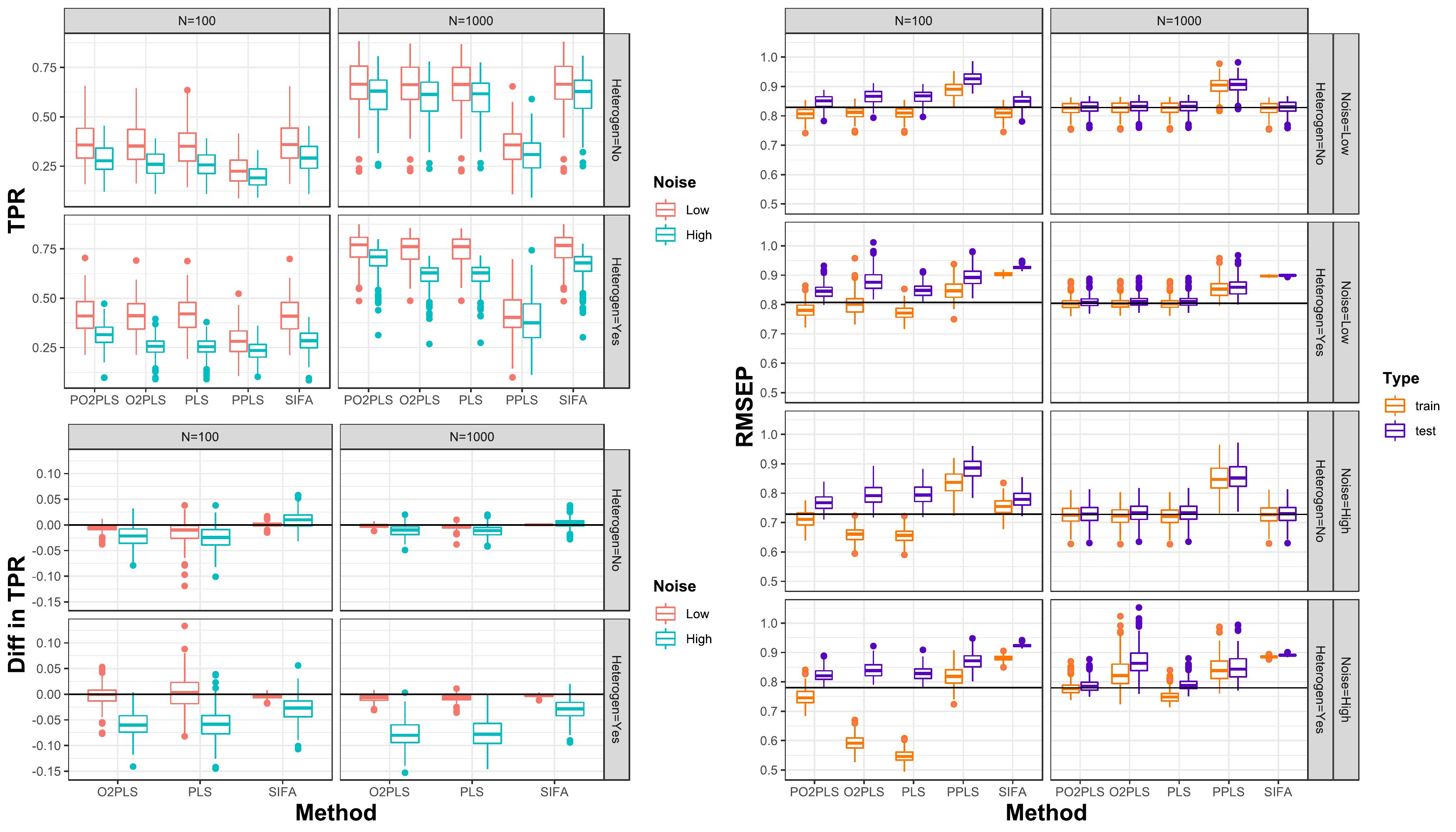}
	\caption{\textbf{Simulation study: feature selection and prediction error.} \textit{Upper left figure}: proportion of true top 25\% among estimated top 25\% (TPR) for each method, stratified by simulation scenario. The left and right boxplots represent low and high noise, respectively. \textit{Lower left figure}: Difference in TPR of several methods and PO2PLS. Lower values are in favor of PO2PLS. \textit{Right figure}: Root mean squared error of prediction stratified by method and scenario. The left and right boxplots represent the training and test error, respectively. The black line represents the median test error when using true parameter values. }
	\label{fig5:simuplot}
\end{figure}

We first present the accuracy and prediction performance in the low dimensional setting, see Figure \ref{fig5:simuplot}. Boxplots of the accuracy and prediction error are shown across the scenarios. Differences in accuracy with respect to PO2PLS are also shown. 
In terms of feature selection, PO2PLS performed good compared to the other methods. When considering the TPR difference between each method and PO2PLS per simulation run, PO2PLS generally had the highest TPR. This difference tend to increase with larger noise proportions and more heterogeneous joint parts settings. The differences between PO2PLS and PPLS are not shown for better visual comparison.
Regarding the prediction error, PO2PLS generally performed better than the other methods. SIFA had the highest prediction error when heterogeneity between the joint parts was present. Furthermore, PLS and O2PLS seemed to overfit in noisy, small sample size scenarios: the training error was lower than the test error compared to the other methods. 
In the high dimensional settings, similar results were obtained. Details can be found in the supplementary material. For the high dimensional settings, the implementation of SIFA  gave `out-of-memory' errors. Hence, we could not include SIFA in these comparisons.

\begin{figure}[ht]
	\includegraphics[width=\textwidth]{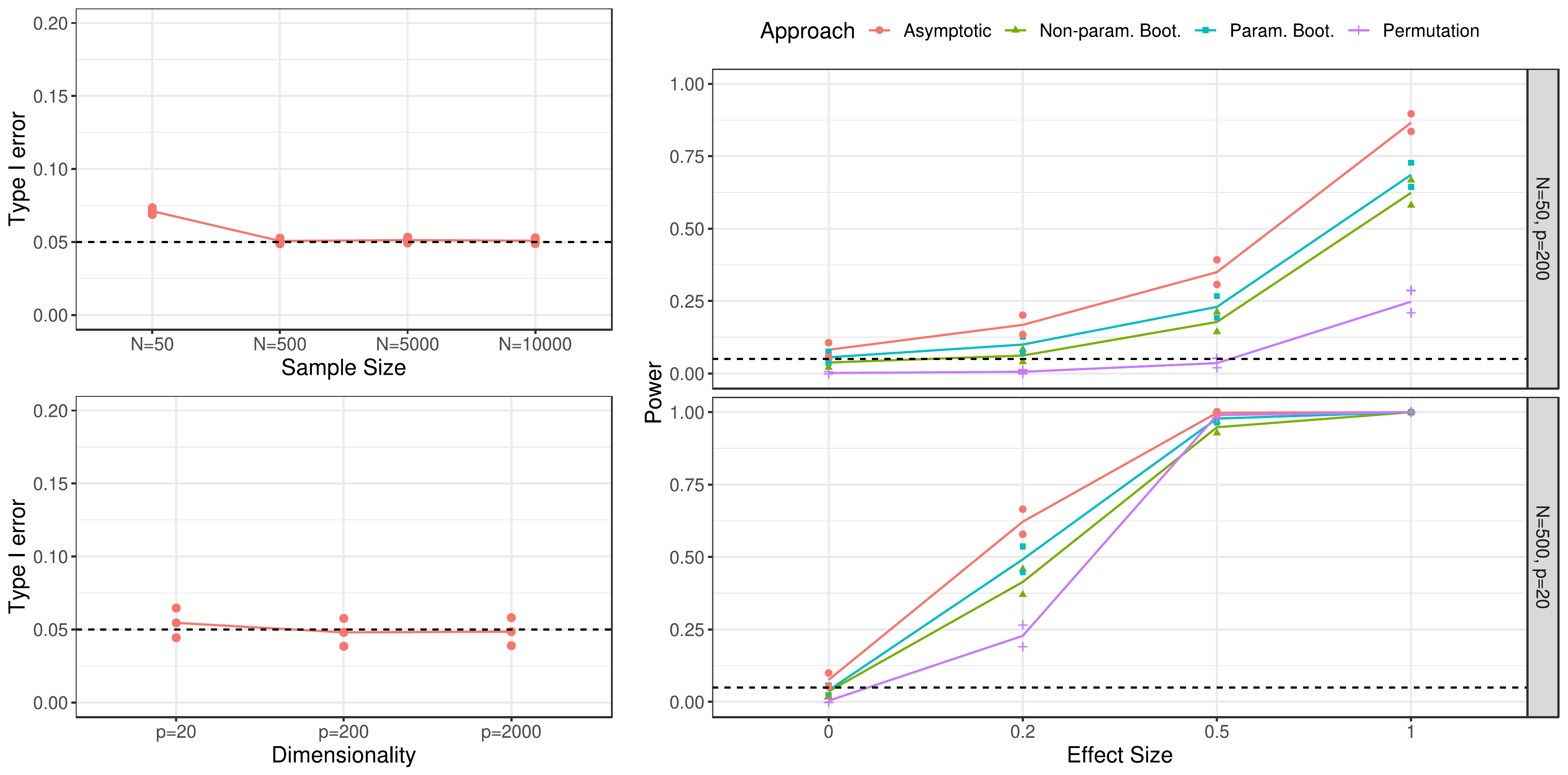}
	\caption{\textbf{Simulation study: global inference.} \textit{Left panel}: Type I error of the global test based on asymptotic PO2PLS statistic, for increasing sample size and $p=20$ (upper plot), and increasing dimensionality and $N=500$ (lower plot). These are based on 50000 and 2000 replicates, respectively. For all plots, 95\% confidence intervals are added based on the binomial distribution, indicated by the points above and below the lines. The dashed horizontal lines represent the 5\% rejection level. \textit{Right panel}: Rejection proportions of the global test performed by the four approaches (asymptotic, non-parametric, parametric bootstrapping, and permutations), for increasing effect size. The sample size was $N=50$ with $p=200$ (upper plot) and $N=500$ with $p=20$ (lower plot). These are based on 500 replicates. }
	\label{fig5:simuH0all}
\end{figure}

{Results for the global inference are shown in Figure \ref{fig5:simuH0all}. The type I error of the PO2PLS test was around 5\% for increasing sample size and dimensionality. Based on the proportion of rejections under the null hypothesis \eqref{eq5:hypothesis}, the PO2PLS test had type I error around 5\% for all but the smallest sample size; in that case, the type I error was about 7\%. It also had more power under the alternative than the other approaches, with the permutation test being severely underpowered in small sample size. }

We briefly present the key results of the additional simulations. In the selected case-control simulation study, PO2PLS had highest TPR, and suffered less from overfitting than PLS and O2PLS. When estimating one component less than the true number of joint and specific components, PO2PLS performed similarly to the algorithmic methods (PLS and O2PLS). Further, PO2PLS was robust against non-normal distributions. Finally, the increase in CPU time and memory usage for increasing data dimensionality was similar across the methods. The full findings are given in the supplementary material.

\clearpage
\section{Applications to omics datasets}
\label{sec5:DataAnalysis}
We illustrate the PO2PLS model with datasets from two different studies. 
Firstly, PO2PLS is applied to test and estimate genetic contributions to glycomic variation in a population based cohort. Secondly, PO2PLS is used to infer a relation between DNA regulation and gene expression using data from a case-control study. Here we also investigate whether the joint components reveal the case control status and whether the top features overlap with findings in cardiovascular diseases. For comparison, we also applied O2PLS to these datasets. 

\subsection{Data integration in a population cohort}
Glycosylation is one of the most common post-translational modifications that enrich the functionality of proteins in many biological processes, such as cell signaling, immune response and apoptosis \citep{Wahl2018}. 
Previously, genome-wide association studies (GWAS) were performed between pairs of single nucleotide polymorphisms (SNPs) and glycans to investigate genetic regulation of glycosylation \citep{Lauc2010, Wahl2018}. 
However, glycans abundances are highly correlated and associated with multiple genes. For example, the glycan G0 was found associated with multiple genes, including \textit{FUT8}, and this gene was itself associated with multiple glycans \citep{Klaric2020}. Therefore a multivariate approach might provide new insights.
%
We first confirm that genetics play a significant role in regulating of glycans. Then, we investigate whether the joint glycan components represent biological structures. Finally, we compare our top genes with genes identified in GWAS.

Genetic and glycomic data were measured, yielding $333858$ genotyped SNPs and $20$ IgG1 glycan abundances for $N=885$ participants in the Croatian Korcula cohort \citep{Lauc2010}. 
%
The SNPs were aggregated on the gene level by combining SNPs around the same gene with PCA, yielding a Genetic PCs (GPCs) dataset.
Then, the GPCs and glycomics datasets were pre-processed, resulting in datasets $X$ ($p=37819$) and $Y$ ($q=20$), respectively. Based on scree plots of the eigenvalues of $X^\top X$, $X^\top Y$ and $Y^\top Y$, five joint, five genetic-specific, and no glycan-specific components were retained.

A global test for the association between genetics and glycans was performed using PO2PLS. The $T_B$ statistic for each component was between four (for the first component) and three (for the last component). With corresponding p-values of $10^{-5}$ and $10^{-3}$, there is statistical evidence of a relationship between genetics and glycans. 

\begin{figure}[ht]
\centering
\includegraphics[width=0.3\textwidth]{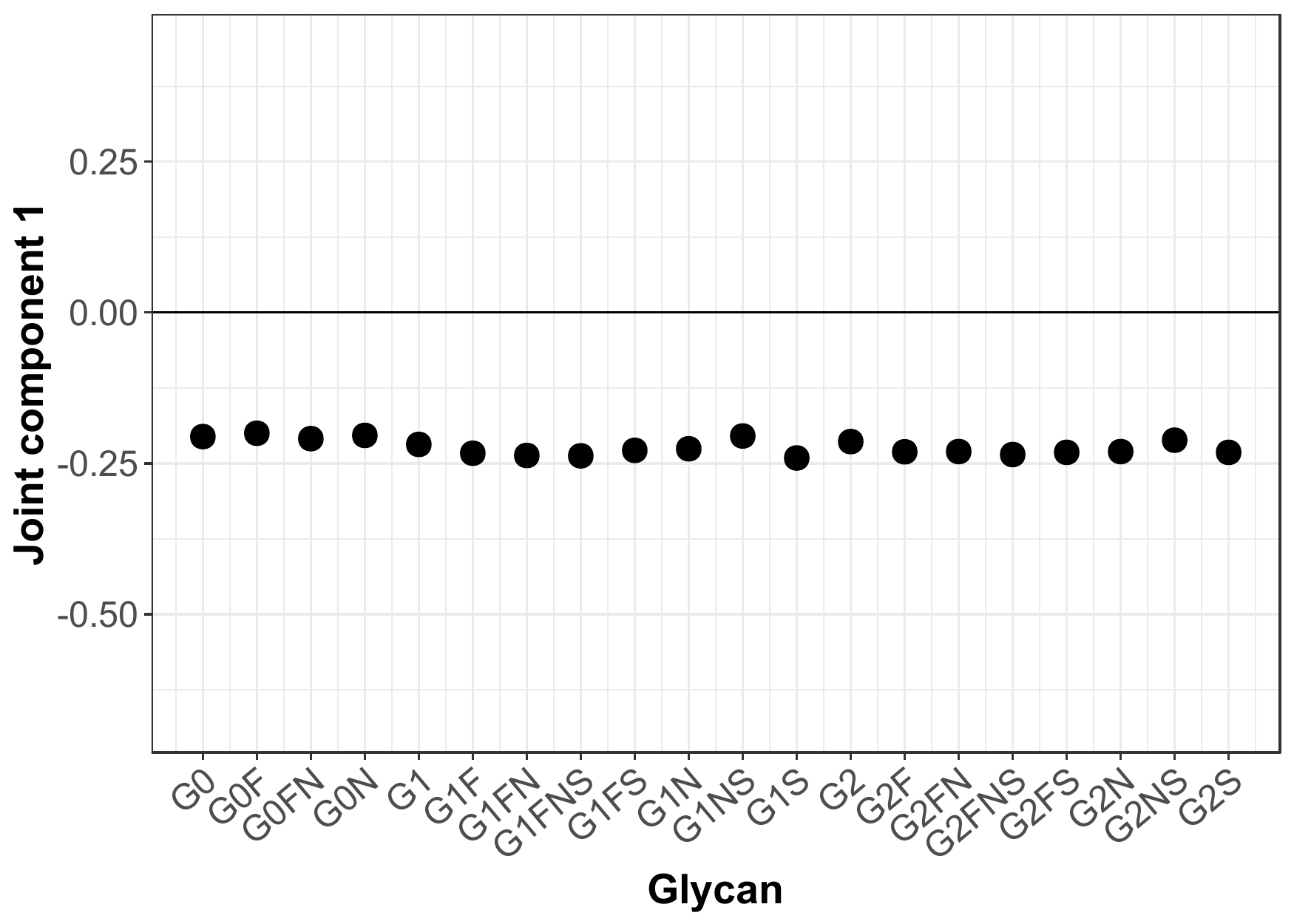} \quad
\includegraphics[width=0.3\textwidth]{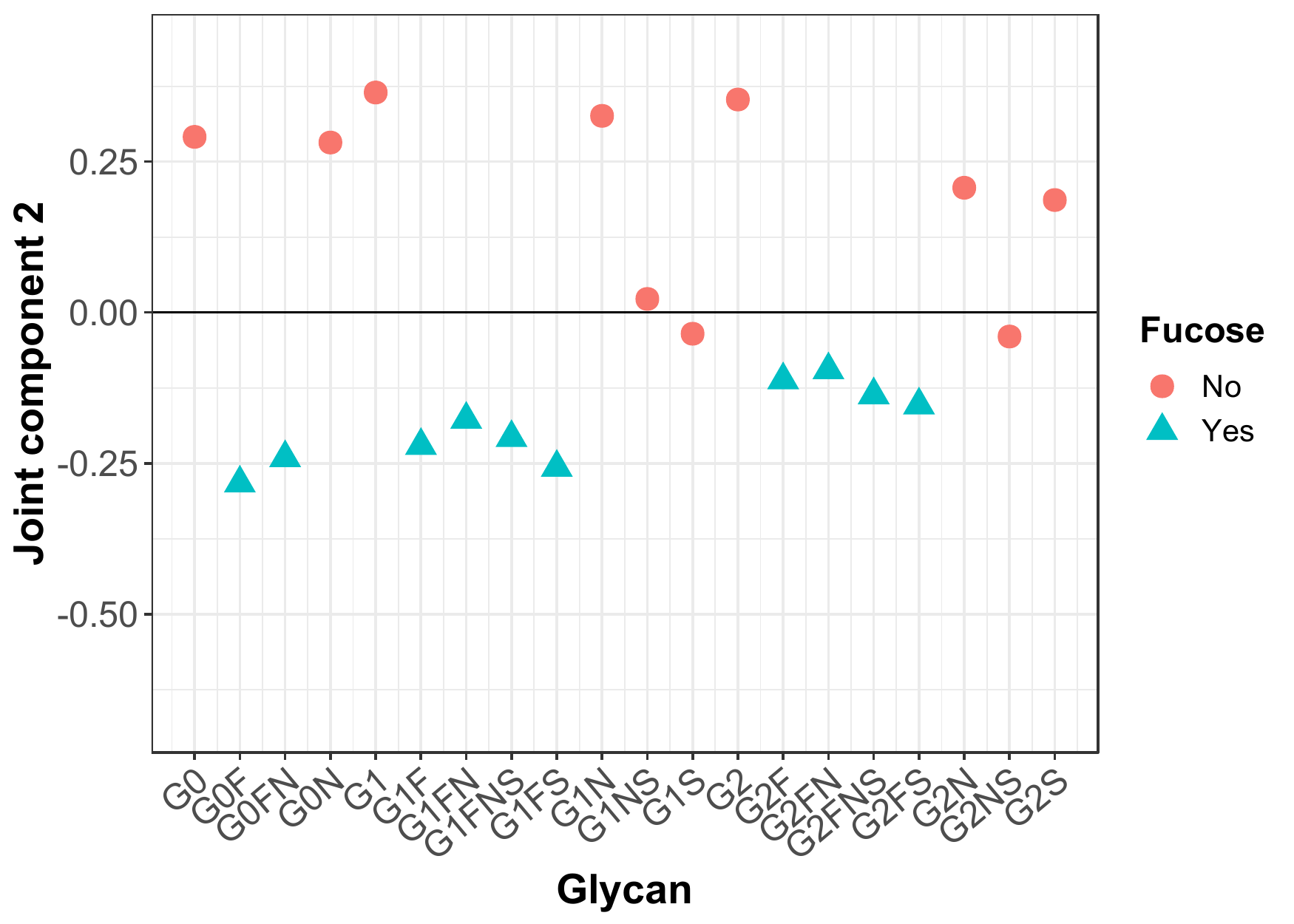} \quad
\includegraphics[width=0.3\textwidth]{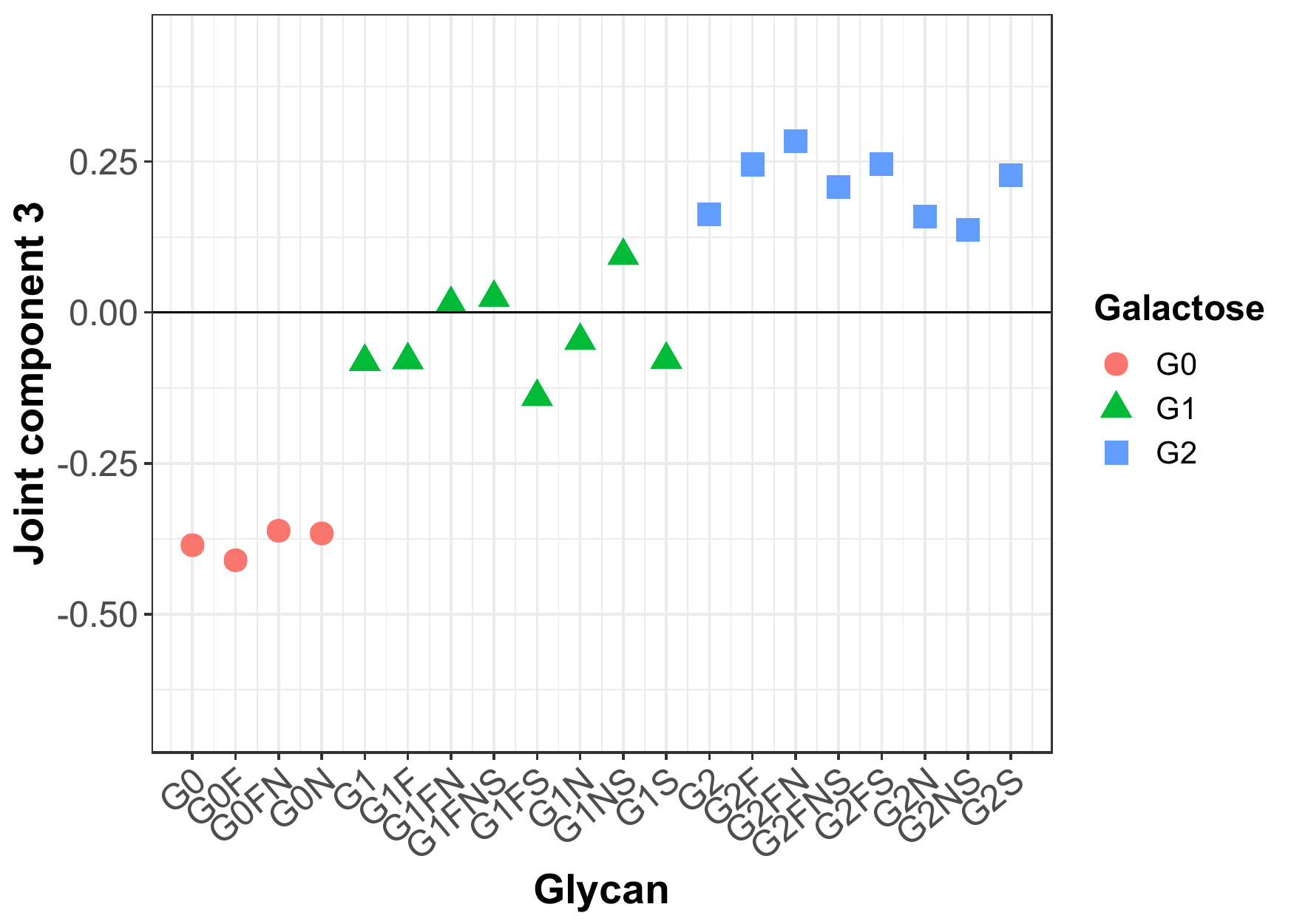} \\
\includegraphics[width=0.3\textwidth]{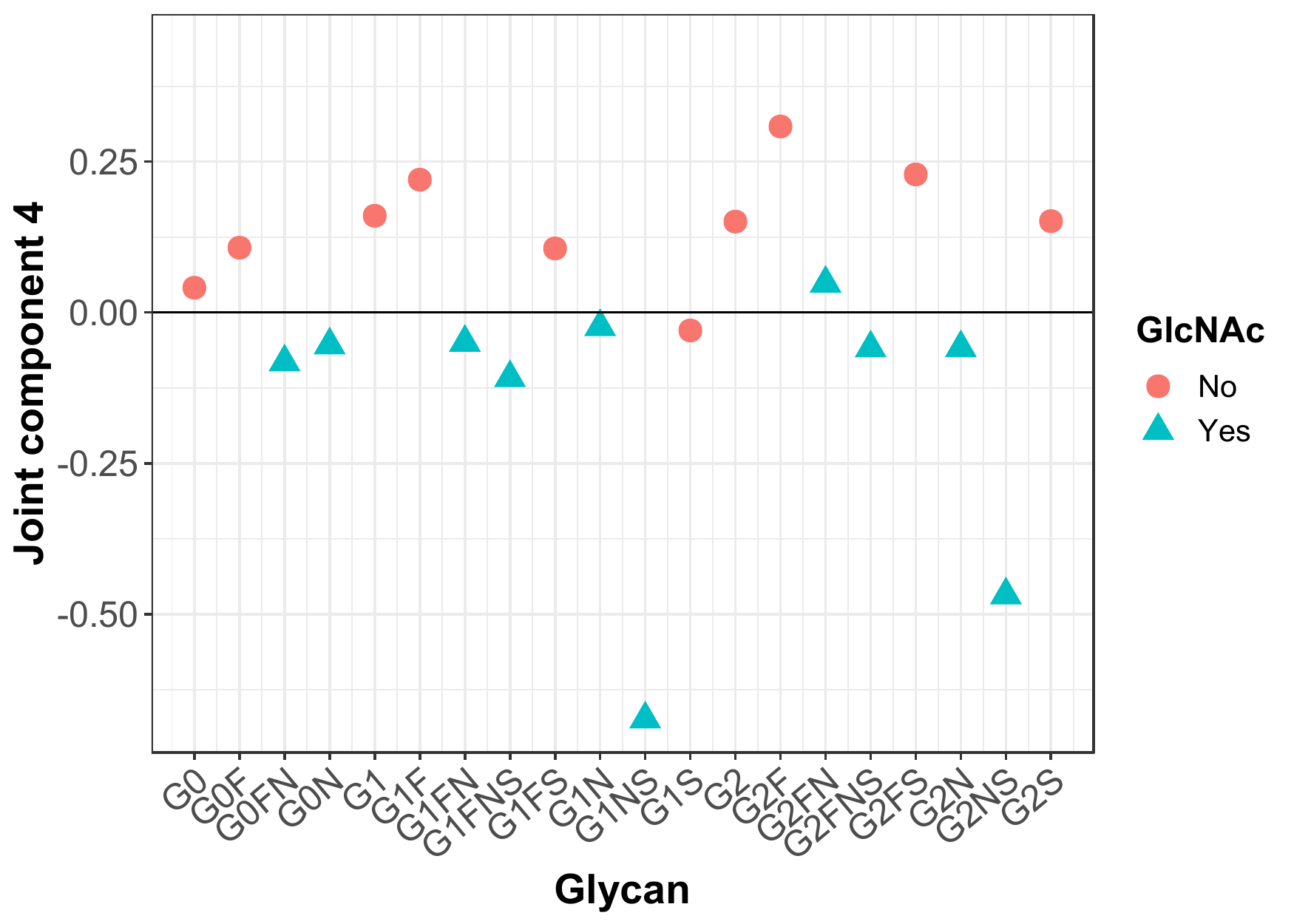} \quad
\includegraphics[width=0.3\textwidth]{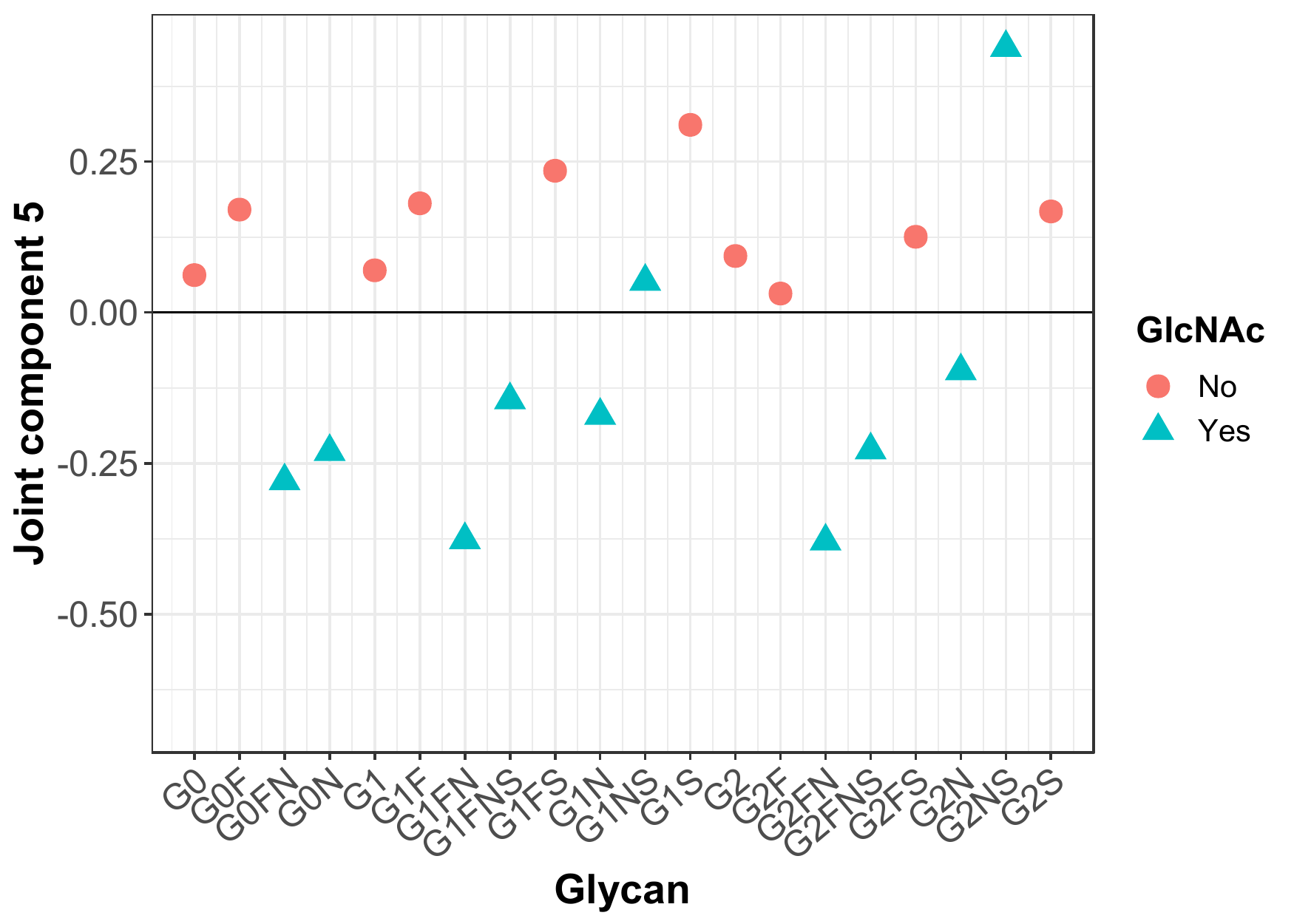} \quad
\includegraphics[width=0.3\textwidth]{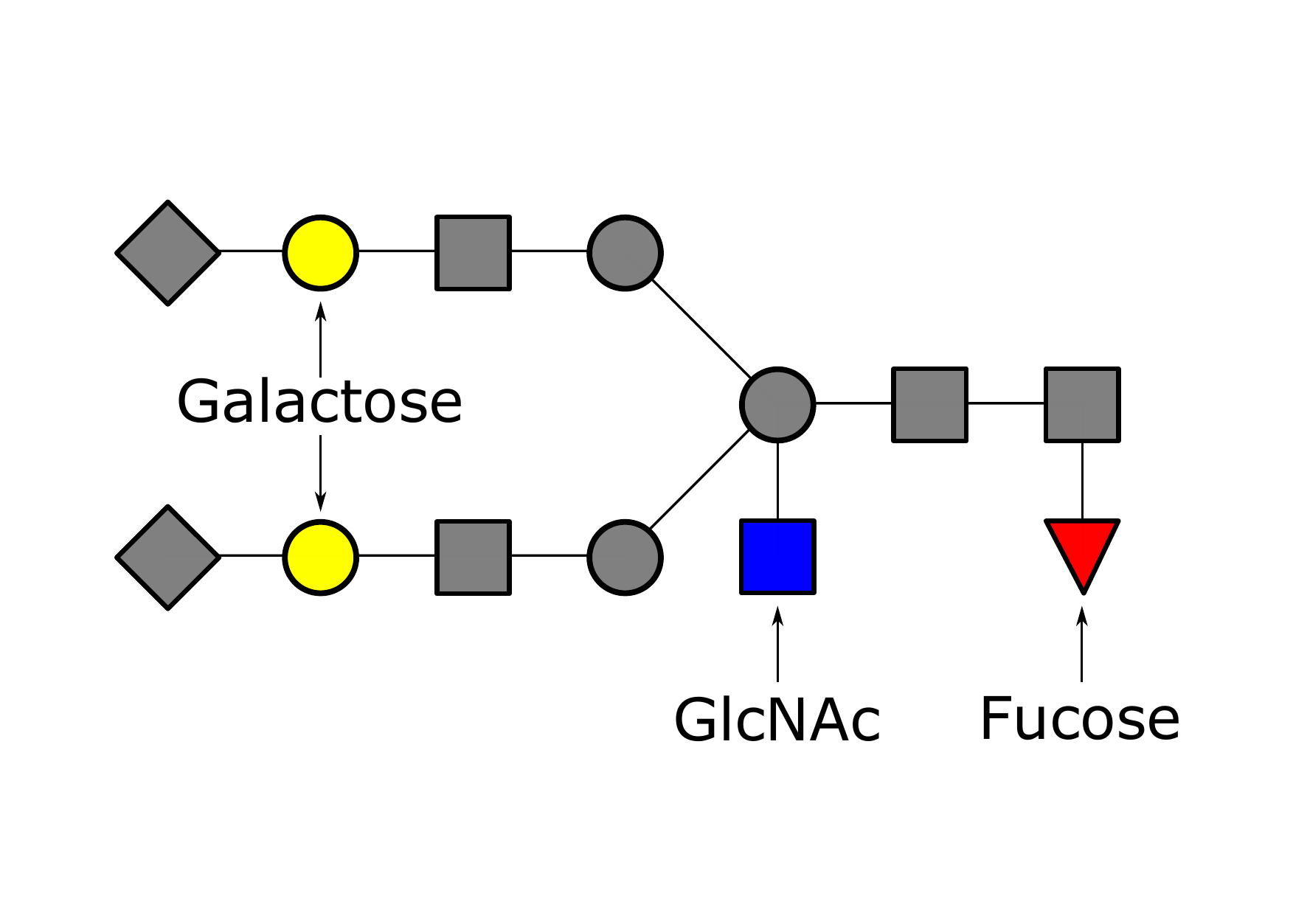}
\caption{\textbf{Glycomic PO2PLS joint components.} The five JPCs are plotted one by one, from left to right, from top to bottom. The dots represent the loading values of each glycan, indicating their importance in the genetic-glycomic relationship. The colors and shapes represent the biological grouping of the glycans. In the last row and column, a graphical representation of the structure of a particular glycan is shown.}
\label{fig5:JointLoadings}
\end{figure}

The loading values of each glycan variable for the five joint components are depicted in Figure \ref{fig5:JointLoadings}. 
Each joint glycan component appears to represent different aspects of glycans and their molecular structure. While the first component represents the `average' glycan (first component), the second component represents presence of fucose, the third component represents the presence of galactose, and the last two components represent GlcNAc \citep{Bouhaddani2018b}.
The top gene in the second joint genetic component is \textit{FUT8} which has been linked to fucosylation \citep{Lauc2010}. Note that the second glycan component reflects ``presence of fucose''. The same article reports more genes linked to glycosylation that we did not find, but their GWAS results are based on imputed genetic data from multiple cohorts. With our joint approach, several other top genes were found, e.g. \textit{DNAJC10} and \textit{AKAP9}, that have links to synthesis and degradation of glycoproteins or (more generally) with inflammation and immune responses. 

A second independent study of $714$ participants from the Croatian Vis cohort is available. To replicate our findings in the Korcula cohort, we apply PO2PLS to this cohort and compare the components underlying the genetics and glycomics data. The results from the second study, shown in the supplementary material, are consistent with the above findings, indicating that the obtained components are not specific to one study.
Finally, we compared the prediction error of $Y$ given $X$ of the models estimated with PO2PLS and O2PLS in Korcula, evaluated using the data from Vis. The ratio of training (Korcula) and test (Vis) error appeared to be 5/23 for O2PLS and 20/21 for PO2PLS. This is conform the simulation study that O2PLS is prone to overfitting.

\subsection{Data integration in a case-control study}

Hypertrophic cardiomyopathy (HCM) is a rare heart muscle disease negatively affecting blood circulation and leading to heart failure. Several studies have shown that several molecular factors, such as epigenetics and gene transcription, play an important role in HCM \citep{Hemerich2019}. 
We investigate whether epigenetic variation affects transcription and test these relationship using PO2PLS. Since the samples consist of HCM cases and controls, an obvious question is whether one of the joint components represents this segregation of cases and controls. 

Data on epigenetics (DNA regulation) and transcriptomics (gene expression) is available, obtained from the heart tissue of thirteen HCM patients and ten controls. Epigenetic data were measured using ChIP-seq, yielding regulation levels of $33642$ regions after pre-processing. Transcriptomics data were measured using RNA-seq, yielding $15882$ expression levels after pre-processing (TMM normalization, followed by log transformation). Statistical challenges are the small sample size of 23 and the large number of features (around $45000$). 

PO2PLS is applied to the epigenetics ($X$) and transcriptomics ($Y$) data, using two joint components and one specific component for both datasets. These numbers are determined using scree plots. The $T_B$ test statistic for the first component was 9.12, and 2.35 for the second component. The p-values were smaller than $0.001$ for the first component and $0.018$ for the second, so the two component were statistically significant.

To investigate whether the top genes in the joint components are involved in cardiovascular outcomes, we clustered the 500 genes with highest loading values in the first joint PC using DisGeNET (a database of gene-disease associations \citep{Sabater-Molina2018}). The top 10 most significant clusters appear to represent a broad spectrum of cardiovascular diseases (Table \ref{tab5:HCM_RNA}.  

\begin{figure}[ht]
	\centering
	\includegraphics[width=\textwidth]{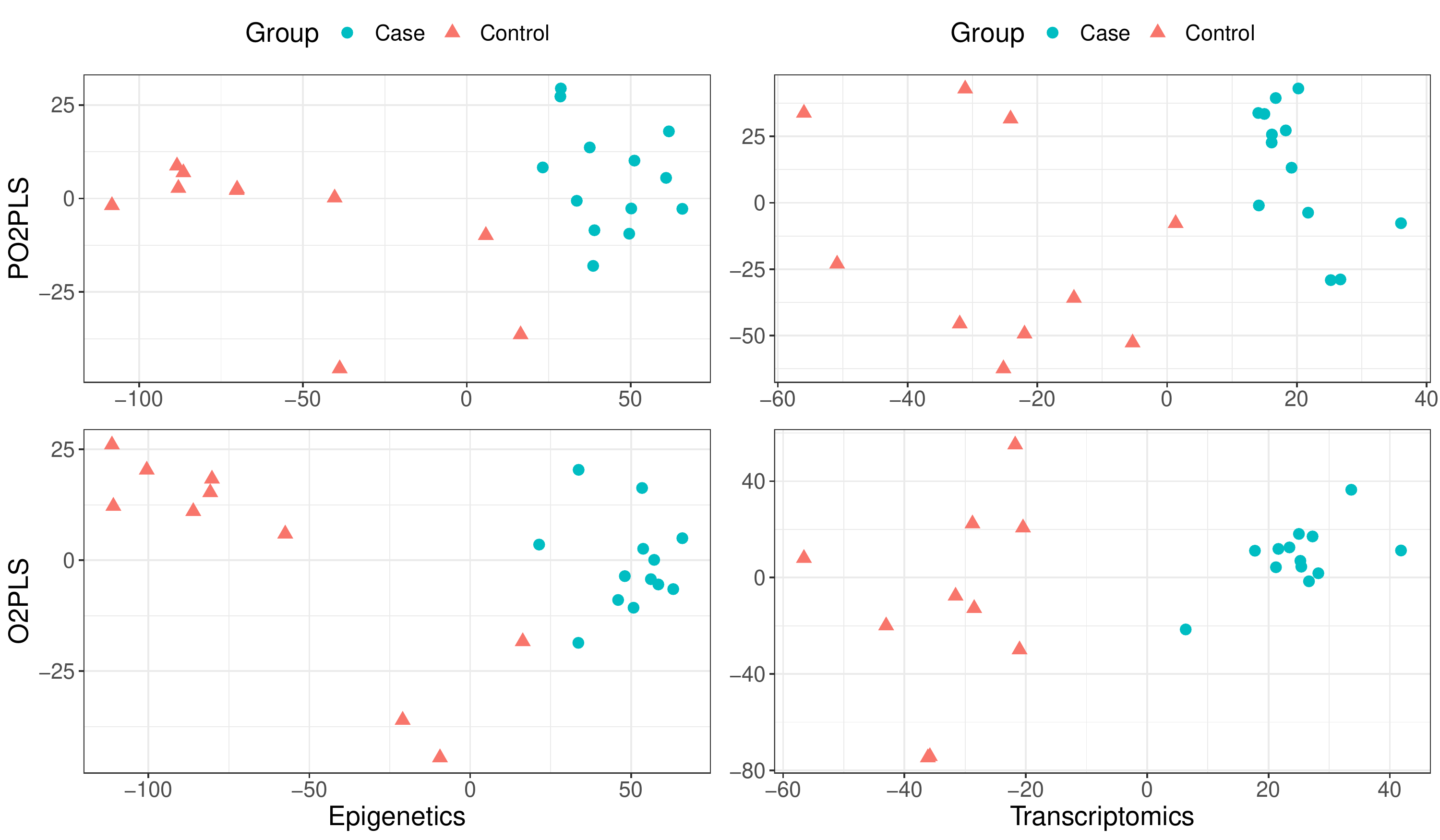} 
	\caption{\textbf{Joint principal component scores across HCM cases and controls.} The two joint component scores are plotted against each other, where the first JPC is on the x-axis. The \textit{upper plots} show transcriptomics resp. epigenetics scores from the PO2PLS fit. The \textit{lower plots} show O2PLS scores. Each dot represents either an HCM patient (blue circle) or a control (red triangle). }
	\label{fig5:ScoresHCM}
\end{figure}

In Figure \ref{fig5:ScoresHCM}, PO2PLS scores are plotted for the first two joint components, and each dot is colored according to its case-control status. The plots indicate that the first joint component picked up the case-control segregation. Additionally, the O2PLS scores are plotted, showing a similar pattern as PO2PLS. 

\begin{table}[ht]
	\caption{\textbf{Annotation of genes in the first transcriptomics joint PC in the HCM analysis.} Using PO2PLS, the top 500 genes were clustered using DisGeNET (a database of gene-disease associations). These top 500 genes are primary drivers of the association with epigenetics across HCM cases and controls. Here, p-values are calculated with a Fisher exact test and corrected for multiple testing. The 10 most significant clusters are shown. }
	\begin{tabularx}{\linewidth}{XXX}
		\textbf{Clusters}	& \textbf{Disease name}		& \textbf{p-value (FDR B\&H)} \\
		\midrule
		Disease cluster 1 	& Hypertensive disease		& 2.53e-7 \\
		Disease cluster 2 	& Arteriosclerosis 			& 2.57e-6 \\
		Disease cluster 3	& Atherosclerosis			& 2.57e-6 \\
		Disease cluster 4 	& Coronary heart disease 	& 7.42e-6 \\
		Disease cluster 5 	& Arthritis 				& 1.19e-5 \\
		Disease cluster 6 	& Aortic Valve Stenosis 	& 2.16e-5 \\
		Disease cluster 7 	& Coronary Artery Disease	& 2.29e-5 \\
		Disease cluster 8 	& Cardiovascular Diseases	& 2.29e-5 \\
		Disease cluster 9 	& Gestational Diabetes 		& 2.67e-5 \\
		Disease cluster 10 	& Heart failure 			& 5.27e-5 \\
		\bottomrule
	\end{tabularx}
	\label{tab5:HCM_RNA}
\end{table}

\clearpage
\section{Discussion}

We propose probabilistic two-way orthogonal partial least squares (PO2PLS) to model the relation between two sets of variables $x$ and $y$ in the presence of data-specific characteristics. Our method is suited for heterogeneous, high dimensional, correlated datasets commonly available in the life sciences. For estimation, we derived a memory efficient EM algorithm. For testing, we derived a Wald type test statistic and its approximate distribution under the null hypothesis of no relationship between $x$ and $y$.  

Via an extensive simulation study, we showed that PO2PLS often performed better than PPLS, SIFA, O2PLS and PLS. In terms of feature selection and prediction, it performed better than PLS, PPLS and SIFA when heterogeneity exists between the datasets. These results were expected since, contrary to the other methods, PO2PLS models the heterogeneity and therefore better estimates the joint components. PO2PLS performed better than O2PLS and PLS in terms of prediction when the datasets are small. For noisy and small datasets, PO2PLS also had a better true positive rate than O2PLS and PLS. PO2PLS had a smaller risk of overfitting, probably because it models all the available information in the data. This reduction in overfitting was also confirmed in studying the relationship between genetic data and glycans, for which we had a replication cohort. The common belief is that PLS and O2PLS, as distribution-free methods, are more suited for small sample size scenarios than probabilistic methods \citep{Wold1985}. Contrary to this belief, in these scenarios, PO2PLS yielded better true positive rate and prediction performance. Via simulations, we also showed that PO2PLS is robust against model deviations such as using a too small number of components and non-normality of the data.

We also showed with simulations that our proposed test statistic for testing the null hypothesis of no relationship is asymptotically normally distributed and performs well in terms of type I error and power. In algorithmic latent variable approaches, testing for a relationship is carried out by empirically estimating the distribution of the test statistic. Since this is time-consuming, evidence for relevance of the top features (e.g. genes, proteins, glycans) is instead obtained by relating the findings to historical ones \citep{Domingo2019}. For example, it is tested whether specific molecular pathways or interaction networks are over-represented in the top feature ranking. Such an approach has the advantage that prior domain knowledge is incorporated. A drawback is a focus on existing findings and a bias against novel discoveries. Moreover, it is often unclear how much evidence exists for pathways and networks in these databases. Each database uses its own scoring mechanisms, often not based on a formal scoring method. Further, there might be a lack of information in the context for new diseases or measurement techniques, and using the information on related diseases or datasets may result in incorrect conclusions about relations \citep{Mubeen2019}. A formal testing procedure quantifies the evidence and might lead to the identification of relevant relationships. 

PO2PLS was applied to omics data from two case studies. The first one is a typical epidemiological population cohort, designed to identify new molecular drivers and build omics predictors for common diseases. These studies are also well suited to study relationships between multiple omics datasets. We applied PO2PLS to genetic and glycomics datasets. The relationship between genetics and glycomics was statistically significant, which confirmed the known high heritability of glycans and the multiple hits of genome-wide association studies (GWAS) \citep{Zaytseva2020}. Moreover, our findings overlapped with GWAS results. We did not replicate all GWAS findings since we restricted ourselves to genotyped SNPs in a gene's neighborhood. On the other hand, modeling the joint distribution of glycans and genes also led to new findings. We replicated the estimated components with relevant features in a second cohort study. The second case study was a small case-control study. To identify molecular markers for rare diseases, omics datasets are measured in cases and controls. Typically these studies are small, either because of the limited number of available cases (rare disease) or costs. Note that multiple diseases can be studied in epidemiological studies, while a case control study is typically limited to one outcome. We applied PO2PLS to epigenetic and transcriptomic data in HCM cases and controls. The relationship between the two sets was statistically significant. Clustering of the top genes using DisGeNET showed that the top genes are in gene clusters associated with several cardiovascular diseases. Moreover, when plotting the first two joint components against each other, a structure representing case control status was evident. This might be expected since all analyses are conditional on the outcome status, and the outcome is a collider for features of the datasets that affect the outcome variable \citep{Balliu2015, Tissier2017}. More research is needed here. 

A possible approach is to include the outcome variable in the model. Several penalized regression models have been used to identify sets of variables related across the different datasets $x$ or $y$ which predict $z$ \citep{Vinga2020}. These approaches do not model the within and across correlations and are hard to interpret when correlations between $x$ and $y$ are present \citep{Tissier2018a}.
For a more holistic approach, one could consider the joint distribution of $(x,y,z)$. Based on the probabilistic O2PLS framework, this distribution can be specified conditional on latent joint and specific variables. In such a framework, the relation between $x$ and $y$ is modeled, and their association with the outcome $z$ is simultaneously incorporated and estimated. Extending our framework in this direction would enable formal tests for the relationship between $x$ and $y$ jointly with the outcome.

More generally, $z$ might be a third dataset instead of an outcome. Here, the interest may lie in inferring relations between the three sets of variables. A complication is that the direction of the relationship between the sets of variables needs to be considered, which might be unknown. The majority of integration approaches for more than two datasets avoid this issue by specifying a common set of latent variables $t$ for all sets of variables \citep{Meng2016}, similar to SIFA. Another approach proposes optimizing a sum of objective functions for each pair of datasets \citep{Lofstedt2011}, while accounting for heterogeneity in the joint parts. 

Many epidemiological cohort studies have multiple omics datasets measured. Currently, we are developing a meta-analysis approach to obtain more robust results by including multiple cohorts in one analysis. For factor analysis, several methods have been proposed to combine the estimated correlation matrices \citep{Cheung2015} or factor loadings \citep{Jak2020} across cohorts. However, the pooling step is not based on the asymptotic variance of the estimators, but an arbitrary covariance matrix. For PO2PLS, the asymptotic variance is available as output (for low-dimensional data). Therefore, as an alternative, the PO2PLS model can be extended by adding cohort-common and cohort-specific parameters to the model. Maximum likelihood estimation would yield an `optimal shared joint space' that incorporates information from each cohort. In such a framework, integration is possible in both `horizontal' (i.e. across studies) and `vertical' (across datasets in the same study) direction. 

Several extensions of the model can be considered. For example, a penalty term can be added to the likelihood function to incorporate prior belief about which variables are more important or belong together. For O2PLS, such a method was recently proposed \citep{Gu2021}. Extending this approach to PO2PLS would be straightforward. Another extension uses functional counterparts to model functional data such as images or temporal data from devices, which are topics of future research. 
To conclude, PO2PLS is a complete framework to test for relationships between omics datasets, identify relevant features and predict outcomes. 

\appendix
\section*{APPENDIX: An EM algorithm for PO2PLS}
\label{ap5:ECMalg}
\begin{AppThm}
	Let $X$ and $Y$ be data matrices with $N$ i.i.d. PO2PLS replicates of $(x,y)$ across the rows. Let $r$, $r_x$ and $r_y$ be fixed, satisfying $\max(r+r_x,r+r_y) < N$. 
	The loading matrix $W$ is estimated with the following iterative scheme in $k$, given known starting values for $k=0$. Here, $\E_k[\cdot] := \E[\cdot | X, Y, \theta^k]$.
	\begin{equation*}
	\begin{split}
	W^{k+1} & = \orth{X^\top \, \E_k\left[ T \right] - \Wo^k \E_k\left[ \Tto^\top T \right]} \\
	\Wo^{k+1}&= \orth{X^\top \, \E_k\left[ \Tto \right] - W^{k+1} \E_k\left[ T^\top \Tto \right]}\\
	C^{k+1} & = \orth{Y^\top \, \E_k\left[ U \right] - \Co^k \E_k\left[ \Uo^\top U \right]} \\
	\Co^{k+1}&= \orth{Y^\top \, \E_k\left[ \Uo \right] - C^{k+1} \E_k\left[ U^\top \Uo \right]} \\
	B^{k+1} & = \mathbb{E}\left[ U^\top T \right] \left(\mathbb{E}\left[ T^\top T \right]\right)^{-1} \circ I_r \\
	\Sigma_{t}^{k+1} & = \frac{1}{N}\E_k\left[ T^\top T \right] \circ I_r \\
	\Sigma_{\tto}^{k+1} & = \frac{1}{N}\E_k\left[ \Tto^\top \Tto \right] \circ I_{r_x} \\
	\Sigma_{\uo}^{k+1} & = \frac{1}{N}\E_k\left[ \Uo^\top \Uo \right] \circ I_{r_y} \\
	\Sigma_{h}^{k+1} & = \frac{1}{N}\E_k\left[ H^\top H \right] \circ I_r \\
	(\sigma^2_{e})^{k+1} & = \frac{1}{Np}\tr\left(\E_k\left[ E^\top E \right]\right) \\
	(\sigma^2_{f})^{k+1} & = \frac{1}{Nq}\tr\left(\E_k\left[ F^\top F \right]\right) \\
	\end{split}
	\end{equation*}
	\label{th5:ECM}
\end{AppThm}
The proof is given in the supplementary material.

\bigskip
\begin{center}
{\large\bf SUPPLEMENTARY MATERIAL}
\end{center}

\begin{description}

\item[Proofs, simulations and details for PO2PLS (pdf):] This document contains additional materials for the methods, simulation and data analysis sections. First, details and proofs of theoretical variances and covariances, identifiability, maximum likelihood estimation and asymptotic results are derived. Then, additional results of the simulation study are shown. Finally, the results of the extra data analysis is shown.

\end{description}

\bibliographystyle{asa2}
\bibliography{library}


\end{document}